\documentclass[conference]{IEEEtran} 

\usepackage{booktabs} 

\usepackage{cite,url,multirow,multicol}

\renewcommand{\arraystretch}{1.1}
\newcommand{\nlcell}[3][1.0]{\def\arraystretch{#1}\begin{tabular}[c]{@{}#2@{}}#3\end{tabular}}

\makeatletter
\def\ps@IEEEtitlepagestyle{%
  \def\@oddfoot{\mycopyrightnotice}%
  \def\@evenfoot{}%
}
\def\mycopyrightnotice{%
  {\footnotesize
  \begin{minipage}{\textwidth}
    \textcopyright 2018 IEEE.  Personal use of this material is permitted.  Permission from IEEE must be obtained for all other uses, in any current or future media, including reprinting/republishing this material for advertising or promotional purposes, creating new collective works, for resale or redistribution to servers or lists, or reuse of any copyrighted component of this work in other works.
  \end{minipage}\hfill}
  \gdef\mycopyrightnotice{}
}

\begin{document}
\title{How angry are your customers? Sentiment analysis of support tickets that escalate}

\author{\IEEEauthorblockN{Colin Werner}
\IEEEauthorblockA{University of Victoria\\
Victoria, BC, Canada\\
colinwerner@uvic.ca}\\
\IEEEauthorblockN{Gabriel Tapuc}
\IEEEauthorblockA{University of Ottawa\\
Ottawa, ON, Canada\\
gtapu088@uottawa.ca}
\and
\IEEEauthorblockN{Lloyd Montgomery}
\IEEEauthorblockA{University of Victoria\\
Victoria, BC, Canada\\
lloydrm@uvic.ca}\\
\IEEEauthorblockN{Diksha Sharma}
\IEEEauthorblockA{University of Victoria\\
Victoria, BC, Canada\\
sharmad@uvic.ca}
\and
\IEEEauthorblockN{Sanja Dodos}
\IEEEauthorblockA{University of Victoria\\
Victoria, BC, Canada\\
sanjadod@uvic.ca}\\
\IEEEauthorblockN{Daniela Damian}
\IEEEauthorblockA{University of Victoria\\
Victoria, BC, Canada\\
danielad@uvic.ca}
}



\maketitle

\begin{abstract}
Software support ticket escalations can be an extremely costly burden for software organizations all over the world. Consequently, there exists an interest in researching how to better enable support analysts to handle such escalations. In order to do so, we need to develop tools to reliably predict if, and when, a support ticket becomes a candidate for escalation. This paper explores the use of sentiment analysis tools on customer-support analyst conversations to find indicators of when a particular support ticket may be escalated. The results of this research indicate a considerable difference in the sentiment between escalated support tickets and non-es\-ca\-la\-ted support tickets. Thus, this preliminary research provides us with the necessary information to further investigate how we can reliably predict support ticket escalations, and subsequently to provide insight to support analysts to better enable them to handle support tickets that may be escalated.
\end{abstract}

\section{Introduction}
Companies selling software products often provide support for their products in the form of person-to-per\-son interaction, whether that be through email, live text chat, or over-the-phone conversations. When a customer contacts a company's support team a support ticket is opened that represents the issue brought forth by this particular customer. A support ticket exists to record all communication between the customer and the support team for information management and historical tracking purposes. Companies, looking to track their customers beyond individual support tickets, may maintain a customer database with various business metrics. These business metrics, such as products purchased, date of product purchases, and average amount purchased per year, are useful for calculating return on investments and the likelihood of product re-purchases. These types of business metrics, although fruitful for a company looking to be successful \cite{businessmetrics}, ignore the sentiment of customers both from a historical and real-time perspective.

In our research, we investigate the sentiment of customers through the conversations they have on their support tickets using leading sentiment analysis tools. A support analyst, provided with the sentiment of a customer, could adapt to a support situation based on the sentiment of the customer. Support ticket escalation is a process offered by some support organizations and occurs when a customer asks for additional help resolving their issue because they feel the current assistance they are receiving is inadequate in some way \cite{ecritsmont}. An escalation can cost a company an unnecessary amount of resources \cite{predsftwrescling}\cite{Sheng:2014:CLD:3056241.3056255}, and may involve a support analyst performing additional non-cus\-tomer-orien\-ted work. In this paper we present analysis that highlights the potential of using sentiment analysis in order to predict an escalation in support tickets. If we are able to reliably and accurately predict escalations then we can help reduce the cost of escalations. However, this research is focused on the \textit{potential}, and not the actual, possibility of predicting an escalation. The actual prediction will be the focus of subsequent research. As such, our research was guided by the following research question:
\begin{description}
\item[RQ:] \hfill \\
Is the sentiment in an escalated support ticket significantly different than the sentiment in a non-es\-ca\-la\-ted support ticket?
\end{description}
The RQ served as a preliminary inquiry to determine if there were statistically significant differences that would be indicative of the phenomenon we were looking for. A lack of any significant differences would have meant that no information could be gained or harnessed towards assisting support analysts. In order to assess our RQ, we focused our efforts on analyzing a proprietary large dataset containing actual support tickets from our large and industrial partner, IBM. The sentiment of the support tickets were extracted using the Watson Natural Language Understanding (NLU) API\footnote{https://www.ibm.com/watson/services/natural-language-understanding/}, SentiStrength\footnote{http://sentistrength.wlv.ac.uk/}, and the Natural Language Toolkit\footnote{https://www.nltk.org/} (NLTK). In addition, the Watson NLU API was used to extract the various emotional readings from the support tickets. The resulting sentimental and emotional datasets were then analyzed and we were able to identify significant concrete differences between the escalated and non-es\-ca\-la\-ted datasets.

This paper is structured as follows: Section \ref{ssec:related} summarizes related work; Section \ref{sec:data} defines how the data was collected; Section \ref{ssec:sent_tools} describes the tools used in this research; Section \ref{ssec:sent_emo_data_results} contains the results of the sentiment analysis; Section \ref{ssec:discussion} discusses the results and meaning of the sentiment analysis; Section \ref{sec:future_work} highlights areas for future work;  Section \ref{sec:threats} provides the threats and limitations of the conclusions; and Section \ref{sec:conclusions} concludes the paper.

\section{Related Work}
\label{ssec:related}
Our previous work performed similar research on IBM support tickets by developing a machine learning classifier to help predict the risk of support ticket escalation\cite{8049142}. Our research was performed in conjunction with multiple stakeholders from IBM and involved an iterative design science methodology to characterize the support process and data to better manage the escalation process\cite{8049142}. We obtained considerable insight into the ability of predicting customer escalations so that to help simplify the job of support analysts industry-wide\cite{8049142}. In this paper, however, we sought to identify the potential value that the sentiment of customer-support analyst conversations, alone, has on our ability to make such predictions when the other information about the support tickets is not available. 

Sentiment analysis is emerging as a popular technology, especially with such a breadth of research topics that are applicable to the sentiment analysis field. As such, there are already a wide variety of sentiment analysis tools available, many of which are open-sou\-rce. Watson NLU, NLTK, SentiStrength, Alchemy\footnote{Acquired by IBM and subsequently deprecated}, Stanford NLP\footnote{\url{https://github.com/stanfordnlp/CoreNLP}}, and EmoTxt\footnote{\url{https://github.com/collab-uniba/Emotion_and_Polarity_SO}} are examples of sentiment analysis tools currently available. The aforementioned list of tools should not be considered exhaustive, as there are a multitude of others that have been developed and even more that are actively being developed. That being said, there are two different varieties of sentiment analysis tools. The first type will classify the sentiment polarity of text; it identifies the text as either neutral, positive, or negative. The second type will analyze text for various emotions and report the findings on either an individual emotion, a range of emotions, or the sentiment of the text. The majority of sentiment analysis tools were trained using openly available content from the internet, including reviews, comments, and social media data, therefore applying sentiment analysis tools to the software engineering domain may not be as useful due to the use of technical jargon\cite{Jongeling:2017:NRU:3135854.3135907}. However, most of the research has been performed on software development artifacts, such as developer commit messages and Stack Overflow. In order to limit the scope of our research, we assumed that the technical jargon commonly seen in software development artifacts will be vastly different from customer support artifacts, as the customer support conversations involve a support analyst and a customer as opposed to interactions between developers. Jongeling et al\cite{7332508} also researched the applicability of a range of sentiment analysis tools in the software engineering domain and found that SentiStrength and NLTK showed the highest degree of correspondence and the highest agreement with each other. Thus, the results of the use of off-the-shelf sentiment analysis tools when applied in the software engineering domain is still actively being researched.

Recent research by Novielli et al\cite{benchmark-novielli} have replicated the study by Jongeling et al\cite{7332508} using three sentiment analysis tools that were unavailable at the time of the original study. These three sentiment analysis tools, Senti4SD\footnote{\url{https://github.com/collab-uniba/Senti4SD}}, SentiStrength-SE\footnote{\url{http://laser.cs.uno.edu/Projects/Projects.html}}, and SentiCR\footnote{\url{https://github.com/senticr/SentiCR}} have been developed specifically for the software engineering domain. The results of this study indicate that customized tools do indeed provide more accurate results. Furthermore, Lin et al\cite{howfar} confirmed with a "negative results" paper warning researchers that the current set of off-the-shelf sentiment analysis tools are not yet mature enough to be effectively used in practical applications in the software engineering domain. However, as previously mentioned our goal is to determine the \textit{potential} usefulness of using sentiment analysis tools in predicting support ticket escalations. As such, we are simply making use of the current off-the-shelf sentiment analysis tools as a stepping stone to provide us with an appropriate level of confidence prior to delving deeper into our research field.

A plethora of sentiment analysis research has already been performed across various domains including the analysis of social media, such as Twitter\cite{Thelwall:2012:SSD:2336247.2336261}\cite{godbole-news-blogs} and other media posted on social networks\cite{Ribeiro2016}. Other studies have researched sentiment analysis on a variety of software engineering domains, including: commit comments in GitHub\cite{Guzman:2014:SAC:2597073.2597118}\cite{Pletea:2014:SES:2597073.2597117}\cite{Sinha:2016:ADS:2901739.2903501}, Gentoo Linux community\cite{6686063}, openSUSE development community\cite{rousinopoulos}, Apache issue reports\cite{Gachechiladze:2017:ADC:3102962.3102966}, GitHub and StackOverflow\cite{Novielli:2015:CSD:2804381.2804387}, artifacts from various sources of JIRA trouble tickets\cite{Ortu:2016:ESS:2901739.2903505}, and other various software development artifacts\cite{Guzman:2013:TEA:2491411.2494578}\cite{Murgia:2014:DFE:2597073.2597086}. There have even been studies analyzing the sentiment of IT support tickets\cite{7832903}; however, this research focused on an internal IT Support organization, whereby both the support analysts and customers were employed by the same organization. The results of the aforementioned IT support ticket research lead to the development of a new sentiment analysis tool that specifically targets IT support tickets with better accuracy than the leading sentiment analysis tools\cite{7832903}. Therefore, a task remains for future research which should involve leveraging this newly developed sentiment analysis tool to corroborate the results discussed in this paper. That being said, all of the related research applied sentiment analysis tools to the software engineering domain and subsequently achieved highly successful results. In summary, we can see that the sentiment analysis tools, while still a maturing and evolving field, can help us provide substantial new research to the software engineering domain.

\section{Data and Processing}
\label{sec:data}
The repository data used in this research consisted of numerous support tickets from IBM.  Each support ticket is an artifact that records a series of emails between a customer and support analysts for a particular case. The workload in processing the support tickets involved splitting each plain-text file into individual emails and then manually labelling each email as either from support to customer (support) or customer to support (customer). The annotation process revealed that some support tickets contained no emails, so these support tickets were subsequently removed. Additionally, some entries within a particular support ticket were not an email and were subsequently removed from the dataset. Details on this dataset, its size and its pre-processing are provided in what follows.  
\subsection{The Data: Support Tickets}
A support ticket is an artifact that records a series of emails between a customer and the various entities of the company's support team regarding  this particular customer's issue. A support ticket exists for the entire duration of a particular customer's issue, which can last several days, weeks, months, or even longer. The emails contained within the support ticket are either sent from a customer to support (customer) or from support to a customer (support). Together, these two types of emails form the entirety of the interactions that will be investigated by this research. A shortcoming of these support tickets is that they do not record or transcribe phone calls, thus leaving a notable information gap when customers prefer over-the-phone support.

At the time of this research, support tickets in IBM's system were stored as plain-text files. This introduced two problems that concerned this research: 1) how to separate emails from each other; and 2) how to classify who sent a particular email, the customer or support. The first problem was solved with a Python parsing script that took advantage of notable patterns in appended emails, which allowed the plain-text of each support ticket to be separated into distinct emails. However, the second problem was not as easily solved due to inconsistencies across support teams as to how emails were documented and stored in the support ticket. To solve this issue, our research team manually annotated the support tickets to classify each of them as being customer or support.
\subsection{Manual Annotation}
\label{ssec:manual_annotation}
The goal of manually annotating each email was to classify each email as either from a customer or from support. However, to further complicate the process of annotating these emails, some system messages attached to the support tickets looked similar to an email, so the annotation process also involved removing these system messages from the dataset. After the manual annotation process was completed, there were 655 support tickets containing a total of 10172 emails.  Once all of the pre-pro\-cess\-ing was complete, the emails could now be analyzed using the sentiment analysis tools.
\subsection{Dividing the Data}
\label{ssec:dividing_data}
The output of the manual annotation in Section \ref{ssec:manual_annotation} included 10172 emails in 655 support tickets; however, the data had to be divided into subsets for the purpose of this research. 
The first division of the data occurs when each email is classified as either a customer email or a support email, as described in Section \ref{ssec:manual_annotation}. This initial division allows the sentiment values that were being tracked during a support ticket to be mutually exclusive as to who is represented: a customer or support.

The second division occurs when the customer and support datasets were further divided into "escalated" and "non-es\-ca\-la\-ted" subsets, thus creating a total of four subsets. In summary, the emails were divided into the following datasets:

\begin{enumerate}
\item customer escalated,
\item customer non-es\-ca\-la\-ted,
\item support escalated, and
\item support non-es\-ca\-la\-ted,
\end{enumerate}
At this point, each email\footnote{Please note that while our aim is to study an entire support ticket entity for the purpose of this paper we used individual emails as the unit of analysis. Section \ref{sec:future_work} discusses this limitation in further detail.}, from each subset, was individually fed into the sentiment analysis APIs, described in Section \ref{ssec:sent_tools}, and the results from each API were then cached for further analysis.
\section{Sentiment Analysis Tooling}
\label{ssec:sent_tools}
For the purpose of this research, three different tools were used in order to perform sentiment analysis on the emails from the support tickets: NLTK, SentiStrength, and Watson NLU. Each of these tools provide sentiment analysis in a similar manner; however, each tool provides its own flavour with respect to the results of the sentiment analysis. This section shall describe each API and how it was used.
\subsection{NLTK API}
\label{ssec:nltkapi}
The NLTK is a popular open-source Python toolkit that also contains multiple sentiment analysis APIs. This research made use of the Sentiment Intensity Analyzer API, which for the scope of this paper shall be referred to as the NLTK API. The NLTK API returns four real values per input, with each value representing the following sentiment of the input text:
\begin{enumerate}
\item positive,
\item neutral,
\item negative, and
\item compound.
\end{enumerate}
The positive (1), neutral (2), and negative (3) values, when summed, will equal exactly 1. The compound (4) value is an aggregate value, formed from the positive, neutral, and negative values, representing the overall sentiment of the input. One can infer the following based on the compound value, \(c\):
\begin{enumerate}
\item \((c < 0) \iff\) negative sentiment,
\item \((c = 0) \iff\) neutral sentiment, and
\item \((c > 0) \iff\) positive sentiment.
\end{enumerate}
However, given the compound value is simply an aggregate of the positive, neutral, and negative values, one can determine the sentiment by first looking at the neutral score. If the neutral score is greater than 0.5, then the sentiment is considered to be neutral; however, if the neutral score is less than or equal to 0.5 then it is considered to be either positive or negative, whichever has the higher value.

The NLTK API was used to analyze every email that was annotated using the process described in Section \ref{ssec:manual_annotation} and the results were tabulated according to the divisions of data described in Section \ref{ssec:dividing_data}. A summary of the results can be found in Section \ref{ssec:sent_emo_data_results}.
\subsection{SentiStrength API}
\label{ssec:sentistrengthapi}
SentiStrength is another popular sentiment analysis tool that has been described and evaluated by a number of peer-reviewed academic articles. Furthermore, it has been used in a large number of research projects. The SentiStrength tool has a large number of features that merit further exploration; however, for the purposes of this research we used the trinary option provided by the SentiStrength API. The trinary option returns three integer values per input, namely:
\begin{enumerate}
\item positive,
\item negative, and
\item overall.
\end{enumerate}
The positive (1) value, \(p\), is assigned an integer value in the range \(1 \leq p \leq 5\), whereby \(p = 1\) indicates a positive sentiment and \(p = 5\) indicates an extremely positive sentiment. Similarly, the negative (2) value, \(n\), is assigned an integer value in the range \(-5 \leq n \leq -1\), whereby \(n = -1\) indicates a negative sentiment and \(n = -5\) indicates an extremely negative sentiment.

The overall (3) value follows an approach described by Thelwall\cite{Thelwall:2012:SSD:2336247.2336261} and is assigned: -1, if \(p + n < 0\); 0, if \(p + n = 0\); or 1, if \(p + n > 0\). A value of: -1 indicates an overall negative sentiment, 0 indicates an overall neutral sentiment, and 1 indicates an overall positive sentiment. However, one caveat is that if \(p + n = 0\) \textit{and} \(p \geq 4\), then the sentiment is considered to be undetermined and the particular input that produced these values is subsequently removed from dataset\cite{7332508}.

The aforementioned SentiStrength API was used for this research to analyze every email that was annotated using the process described in Section \ref{ssec:manual_annotation} and the results were tabulated according to the divisions of data described in Section \ref{ssec:dividing_data}. A summary of the results can be found in Section \ref{ssec:sent_emo_data_results}.
\subsection{Watson API}
\label{ssec:watsonapi}
For this research we utilized the Watson NLU API. This API offers nine distinct natural language features, two of which are used in this research: emotion and sentiment. To use the NLU API, plain-text is sent to the API with either the sentiment or emotion feature requested and the appropriate metrics are returned. The two NLU APIs utilized for this research each return unique and distinct metrics. 

The NLU Emotion API returns the level of each of the five basic emotions\cite{ekman1972emotion}, sadness, joy, fear, disgust, and anger,  that are measured from the plain-text input. Each emotion is assessed a real value between 0 and 1; whereby 0 indicates the emotion is not being exhibited and 1 indicates a high level of confidence that the emotion is being exhibited.

The NLU Sentiment API returns the level of sentiment being expressed in the  plain-text input. "Sentiment" is a view, attitude, or emotion that is expressed by the words chosen in a sentence \cite{Liu10sentimentanalysis}. The NLU sentiment feature returns a real value between -1 and 1, where -1 indicates an overly negative sentiment and 1 indicates an overly positive sentiment.

As with SentiStrength and NLTK, we used the emotion and sentiment APIs from Watson NLU to analyze every email and the results were tabulated according to the divisions of data described in Section \ref{ssec:dividing_data}. A summary of the results can be found in Section \ref{ssec:sent_emo_data_results}.

\section{Sentimental \& Emotional Data Results}
\label{ssec:sent_emo_data_results}
To address our RQ, we analyzed the customer escalated, customer non-es\-ca\-la\-ted, support escalated, and support non-es\-ca\-la\-ted datasets using three sentiment analysis tools: NLTK, Sentistrength, and Watson NLU. After compiling the data from the tools, we first computed the average of each tool's output for all datasets. In addition, the difference between each average was calculated. Respectively, the averages and differences of the datasets can be found for NLTK, SentiStrength, and Watson NLU in Tables \ref{tab:nltk_avg}, \ref{tab:sentistrength_avg}, and \ref{tab:wat_avg}. Further discussion of these results shall appear in Section \ref{ssec:discussion}.

\begin{table}
\caption{NLTK - Averages}
\label{tab:nltk_avg}
\begin{tabular}{l l l l}
    \hline
    \textbf{Sentiment} & \textbf{Escalated} 
    & \textbf{Non-Escalated} & \textbf{\%-age Change} \\
    \hline\hline
        \multicolumn{4}{c}{Customer Analysis}
    \\\hline\hline
        Compound & 0.278301 & 0.387733 & 39\%
    \\\hline
        Positive & 0.089594 & 0.106865 & 19\%
    \\\hline
        Neutral & 0.871785 & 0.858120 & -2\%
    \\\hline
        Negative & 0.032771 & 0.035022 & 7\%
    \\\hline\hline
        \multicolumn{4}{c}{Support Analyst Analysis}
    \\\hline\hline
        Compound & 0.457937 & 0.544874 & 19\%
    \\\hline
        Positive & 0.102066 & 0.114807 & 12\%
    \\\hline
        Neutral & 0.872724 & 0.857422 & -2\%
    \\\hline
        Negative & 0.024706 & 0.026055 & 5\%
    \\\hline
\end{tabular}
\end{table}

\begin{table}
\caption{SentiStrength - Averages}
\label{tab:sentistrength_avg}
\begin{tabular}{l l l l}
    \hline
    \textbf{Sentiment} & \textbf{Escalated} 
    & \textbf{Non-Escalated} & \textbf{\%-age Change} \\
    \hline\hline
        \multicolumn{4}{c}{Customer Analysis}
    \\\hline\hline
        Overall & 0.080222 & 0.200000 & 149\%
    \\\hline
        Positive & 1.964517 & 2.230769 & 14\%
    \\\hline
        Negative & -1.619253 & -1.683077 & 4\%
    \\\hline\hline
        \multicolumn{4}{c}{Support Analyst Analysis}
    \\\hline\hline
        Overall & 0.237968 & 0.414802 & 74\%
    \\\hline
        Positive & 2.242132 & 2.435456 & 9\%
    \\\hline
        Negative & -1.647794 & -1.614458 & -2\%
    \\\hline
\end{tabular}
\end{table}

\begin{table}
\caption{Watson NLU - Averages}
\label{tab:wat_avg}
\begin{tabular}{l l l l}
    \hline
    \textbf{Emotion} & \textbf{Escalated} & \textbf{Non-Escalated}
    & \textbf{\%-age Change} \\
    \hline\hline
        \multicolumn{4}{c}{Customer Analysis}
    \\\hline\hline
        Anger & 0.074828 & 0.070261 & -6\%
    \\\hline
        Disgust & 0.042419 & 0.031909 & -25\%
    \\\hline
        Fear & 0.060669 & 0.057919 & -5\%
    \\\hline
        Joy & 0.166685 & 0.184073 &10\%
    \\\hline
        Sadness & 0.208549 & 0.206040 & -1\%
    \\\hline
        Sentiment & 0.120708 & 0.194327 & 61\%
    \\\hline\hline
        \multicolumn{4}{c}{Support Analyst Analysis}
    \\\hline\hline
        Anger & 0.063902 & 0.059661 & -7\%
    \\\hline
        Disgust & 0.034842 & 0.032092 & -8\%
    \\\hline
        Fear & 0.055817 & 0.054628 & -2\%
    \\\hline
        Joy & 0.187015 & 0.199418 & 7\%
    \\\hline
        Sadness & 0.185572 & 0.163391 & -12\%
    \\\hline
        Sentiment & 0.258169 & 0.364316 & 41\%
    \\\hline
\end{tabular}
\end{table}


\subsection{Normality Testing}
\label{ssec:normality}
In choosing which difference test to use, the normality of the data contributes to that decision. Normal data allow for stronger conclusions to be drawn using more powerful tests \cite{D'Agostino:1986:GT:19293}. The results of applying the D'Agostino-Pearson\cite{d1973tests} normality test to our data can be found in Tables \ref{tab:nltk_norm_diff}, \ref{tab:senti_norm_diff}, and \ref{tab:e_norm_diff} for each of the sentiment analysis tools. The results of the D'Agostino-Pearson normality test indicate our data were mostly non-normal, as the resulting p-values were considerably less than 0.05. 



\begin{table}[t]
\caption{NLTK - Normality \& Difference Testing}
\label{tab:nltk_norm_diff}
\begin{tabular}{l l l l}
    \hline
        \multirow{2}{*}{\textbf{Sentiment}}
        & \multicolumn{2}{|c|}{\nlcell{c}{\textbf{Pearson P-Value}}}
        & \multirow{2}{*}{\nlcell{c}{\textbf{Mann-Whitney}\\ \textbf{2-Tailed P-Value}}}
    \\\cline{2-3}
        & \multicolumn{1}{|c|}{\textbf{Escalated}} & \multicolumn{1}{|c|}{\textbf{Non-Escalated}} &
    \\\hline\hline
        \multicolumn{4}{c}{Customer Analysis}
    \\\hline\hline
        Compound & 0.000000 & 0.000000 & 0.000012 \quad \(< 0.05\)
    \\\hline
        Positive & 0.000000 & 0.000000 & 0.000002 \quad \(< 0.05\) 
    \\\hline
        Neutral & 0.000000 & 0.000000 & 0.000147 \quad \(< 0.05\) 
    \\\hline
        Negative & 0.000000 & 0.000000 & 0.036273 \quad \(< 0.05\) 
    \\\hline\hline
        \multicolumn{4}{c}{Support Analyst Analysis}
    \\\hline\hline
        Compound & 0.000000 & 0.000000 & 0.001383 \quad \(< 0.05\)
    \\\hline
        Positive & 0.000000 & 0.000000 & 0.000004 \quad \(< 0.05\) 
    \\\hline
        Neutral & 0.000000 & 0.000000 & 0.000015 \quad \(< 0.05\) 
    \\\hline
        Negative & 0.000000 & 0.000000 & 0.176728
    \\\hline
\end{tabular}
\end{table}

\begin{table}[t]
\caption{SentiStrength - Normality \& Difference Testing}
\label{tab:senti_norm_diff}
\begin{tabular}{l l l l}
    \hline
        \multirow{2}{*}{\textbf{Sentiment}}
        & \multicolumn{2}{|c|}{\nlcell{c}{\textbf{Pearson P-Value}}}
        & \multirow{2}{*}{\nlcell{c}{\textbf{Mann-Whitney}\\ \textbf{2-Tailed P-Value}}}
    \\\cline{2-3}
        & \multicolumn{1}{|c|}{\textbf{Escalated}} & \multicolumn{1}{|c|}{\textbf{Non-Escalated}} &
    \\\hline\hline
        \multicolumn{4}{c}{Customer Analysis}
    \\\hline\hline
        Overall & 0.000000 & 0.000000 & 0.008909 \quad \(< 0.05\)
    \\\hline
        Positive & 0.000000 & 0.003232 & 0.000001 \quad \(< 0.05\) 
    \\\hline
        Negative & 0.000000 & 0.000000 & 0.051594 
    \\\hline\hline
        \multicolumn{4}{c}{Support Analyst Analysis}
    \\\hline\hline
        Overall & 0.000000 & 0.000000 & 0.000004 \quad \(< 0.05\)
    \\\hline
        Positive & 0.000000 & 0.000010 & 0.000005 \quad \(< 0.05\) 
    \\\hline
        Negative & 0.000000 & 0.000000 & 0.251714
    \\\hline
\end{tabular}
\end{table}

\begin{table}[t]
\caption{Watson NLU - Normality \& Difference Testing}
\label{tab:e_norm_diff}
\begin{tabular}{l l l l}
    \hline
        \multirow{2}{*}{\textbf{Emotion}}
        & \multicolumn{2}{|c|}{\nlcell{c}{\textbf{Pearson P-Value}}}
        & \multirow{2}{*}{\nlcell{c}{\textbf{Mann-Whitney}\\ \textbf{2-Tailed P-Value}}}
    \\\cline{2-3}
        & \multicolumn{1}{|c|}{\textbf{Escalated}} & \multicolumn{1}{|c|}{\textbf{Non-Escalated}} &
    \\\hline\hline
        \multicolumn{4}{c}{Customer Analysis}
    \\\hline\hline
        Anger & 0.000000 & 0.000000 & 0.017498 \quad \(< 0.05\)
    \\\hline
        Disgust & 0.000000 & 0.000000 & 0.002044 \quad \(< 0.05\) 
    \\\hline
        Fear & 0.000000 & 0.000000 & 0.117554
    \\\hline
        Joy & 0.000000 & 0.000000 & 0.828245
    \\\hline
        Sadness & 0.000000 & 0.000000 & 0.646006
    \\\hline
        Sentiment & 0.000000 & 0.000982 & 0.006840 \quad \(< 0.05\) 
    \\\hline\hline
        \multicolumn{4}{c}{Support Analyst Analysis}
    \\\hline\hline
        Anger & 0.000000 & 0.000000 & 0.391951
    \\\hline
        Disgust & 0.000000 & 0.000000 & 0.540574
    \\\hline
        Fear & 0.000000 & 0.000000 & 0.434205
    \\\hline
        Joy & 0.000000 & 0.000000 & 0.856163
    \\\hline
        Sadness & 0.000000 & 0.000000 & 0.003692 \quad \(< 0.05\) 
    \\\hline
        Sentiment & 0.000000 & 0.000000 & 0.000000 \quad \(< 0.05\) 
    \\\hline
\end{tabular}
\end{table}
\subsection{Difference Testing with Mann-Whitney}
To further address our RQ, difference testing was applied to all datasets comparing each escalated dataset to the corresponding non-es\-ca\-la\-ted dataset. Due to the non-nor\-mal\-ity of our datasets, shown in the Section \ref{ssec:normality}, the Mann-Whitney\cite{mann1947} difference test was chosen using 2-tailed p-value thresholds. The results of the difference testing revealed that there are statistically significant differences between the escalated and non-es\-ca\-la\-ted datasets across all three of the sentiment analysis tools used for this research.

\section{Discussion}
\label{ssec:discussion}
This Section provides a discussion and further analysis of the data from Section \ref{ssec:sent_emo_data_results}.

\subsection{NLTK}
The NLTK results in Table \ref{tab:nltk_avg} show considerable differences between the escalated and non-es\-ca\-la\-ted datasets. The biggest difference can been seen by looking at the average of the customer compound sentiment value for escalated and non-es\-ca\-la\-ted datasets. In particular, the customer compound sentiment value for non-es\-ca\-la\-ted is 39\% higher than escalated, which is a significant difference. Furthermore, the customer positive sentiment for non-es\-ca\-la\-ted is 19\% higher than escalated. Interestingly, the customer negative sentiment value is actually lower for escalated than non-es\-ca\-la\-ted, while the customer neutral sentiment is only slightly higher. The overall averages for support (compound, positive, neutral, and negative) datasets share similar statistical trends, albeit with slightly less magnitude. Moreover, the difference testing in Table \ref{tab:nltk_norm_diff} shows, with high confidence (\(p < 0.001\)), that the escalated and non-es\-ca\-la\-ted datasets are indeed different.  The exceptions include the customer negative sentiment (\(p = 0.036273\)), support compound sentiment (\(p = 0.001383\)), and support negative sentiment (\(p = 0.176728\)) datasets. However, two of aforementioned exceptions maintain \(p < 0.05\), thus our confidence remains high. These statistically significant differences provide supporting evidence to help answer our RQ by showing that the escalated and non-es\-ca\-la\-ted datasets are identifiably distinct. 

\subsection{SentiStrength}
The analysis of the sentiment data provided by SentiStrength in Table \ref{tab:sentistrength_avg} shows a remarkable difference between the escalated and non-es\-ca\-la\-ted datasets. For the customer datasets, the overall sentiment for the non-es\-ca\-la\-ted is 149\% higher than escalated. Similarly, for the support datasets, the overall sentiment for the non-es\-ca\-la\-ted is 75\% higher than escalated. From an arithmetic mean point of view, these two sets are clearly distinct. Furthermore, the difference testing in Table \ref{tab:senti_norm_diff} shows, with considerable confidence (\(p < 0.05\)), that there are statistically significant differences between the escalated and non-es\-ca\-la\-ted datasets for the overall and positive values for both customer and support datasets. In fact, the customer positive, support overall, and support positive all maintain \(p < 0.00001\), while customer overall is slightly higher with \(p < 0.01\). However, the same conclusion cannot be applied to the negative sentiment datasets for both customer (\(p = 0.051594\)) and support (\(p = 0.251714\)).

Nonetheless, the results from SentiStrength corroborate our findings from the NLTK results. As such, our confidence in answering our RQ has increased, as we have shown via two independent tools that there are strong statistical indications that the escalated and non-es\-ca\-la\-ted datasets are different.

\subsection{Watson NLU}
Analysis of the average Watson NLU emotions and sentiment in Table \ref{tab:wat_avg} show some interesting details. First, the customer disgust for non-es\-ca\-la\-ted is 25\% less than escalated, while the customer joy is 10\% more. Second, the overall customer sentiment is 61\% higher for non-es\-ca\-la\-ted versus escalated. This exceptionally large difference for sentiment between escalated and non-es\-ca\-la\-ted is a very telling indication that there are considerable differences between the two datasets. Third, the support sadness is 12\% lower for non-es\-ca\-la\-ted versus the escalated. Finally, the support sentiment is 41\% greater for non-es\-ca\-la\-ted. These statistics provide additional evidence in our continuing confidence to answer our RQ. One can speculate that not only can we identify differences between the escalated and non-escalated datasets, but we can observe that customers who have a support ticket escalated display more disgust, less joy, and a lower overall sentiment than those that are not escalated and support analysts dealing with escalated support tickets display more sadness and a lower overall sentiment. The details of why these particular emotions are more prominent are potential areas for future research, but further speculation can also be left to the reader's imagination. In addition, the Mann-Whitney tests in Table \ref{tab:e_norm_diff} reveal that five of the twelve subsets have a significant difference between them, with p-values less than 0.05. The customer subset has three emotions, anger, disgust, and sentiment, each of which have significant differences. The support subset has two emotions, sadness and sentiment, that have significant differences. These results highlight the importance of using difference testing in addition to standard arithmetic mean, as it can reveal otherwise undiscoverable information in research such as this.

\subsection{Answering Our RQ}
Based on the results of the three independent sentiment analysis tools, we have shown significant statistical differences between the escalated and non-es\-ca\-la\-ted datasets. Thus, we can confidently provide a concrete answer to our RQ: the sentiment in an escalated support ticket is indeed significantly different than the sentiment in a non-es\-ca\-la\-ted support ticket. We can then turn our focus to the \textit{implications} of exactly how we can use this research to further advance our understanding of predicting if, and when, a support ticket may be escalated.

Previous work by Blaz and Becker\cite{7832903} developed a tool that classifies the polarity of sentiment in IT support tickets. However, this tool is not yet perfect, as it continues to misclassify neutral sentiment as negative sentiment. Thus, the maturity of using sentiment analysis tools in the support ticket domain is an immature field and certainly merits future work. Our research provides additional novel contributions in two directions. First, we provide positive results of how off-the-shelf sentiment analysis tools \textit{can} be used in the software engineering domain. Second, we have shown that there exists the \textit{potential} in using  sentiment analysis to predict the escalation of a support ticket. Our preliminary research simply demonstrates the motivation and justification for additional research into methods that use sentiment analysis to differentiate and predict support tickets at high risk of escalation.

In particular, it should be noted that each of the metrics utilized in this research are document-level assessments that calculate values based on the entire plain-text sent to the API. For the purpose of this research, each "document" is an email that is processed. Therefore, after being processed by the sentiment analysis tools each email is assigned a particular set of metrics. However, it is important to note that the overall goal of our research is to analyze a particular support ticket and not individual emails. Therefore, some sort of averaging mechanism is required to summarize each support ticket, since any single support ticket could have contained multiple emails. Therefore, further research is required in order to provide analysis at a support ticket level. In addition, this preliminary research was performed in order to assess whether sentiment analysis tools could be used in order to predict the escalation of a particular support ticket. Thus, further research is required in order to develop a model that will enable support analysts to predict, in real-time, as to when a particular support ticket may be escalated. We discuss these two topics in further detail in the next section.
\section{Future Work}
\label{sec:future_work}
Previous research has shown that leading sentiment analysis tools provide decent results in the software engineering domain; although, it has also shown that the results are not highly accurate\cite{7332508}. That being said, the sentiment analysis tools used in our research have provided us with sufficient evidence in order to confidently answer our RQ and have thus served their purpose. In order to provide further supporting evidence to our RQ we need to perform additional sentiment analysis using tools that were developed specifically for the software engineering domain, IT support tickets in particular, as described in other relevant works\cite{7832903}\cite{Gachechiladze:2017:ADC:3102962.3102966}\cite{benchmark-novielli}.

As previously mentioned, this paper focused on the sentiment of individual emails across the datasets. However, this paper does not examine the sentiment of a particular support ticket, which may contain multiple emails. Therefore, we propose an additional research question: Are the \textit{trends} in the sentiment of an escalated support ticket significantly different than the \textit{trends} in the sentiment of a non-es\-ca\-la\-ted support ticket? In order to answer this research question we must further develop a solution that will enable us to analyze \textit{how} the sentiment \textit{change} over the lifespan of a particular support ticket, which we shall call the sentimental tendency of a support ticket. The tendency value shall take into consideration the highs, lows, and trends of sentiment across all emails in a particular support ticket to attempt to capture a certain subset of sentimental situations. In conjunction, a similar research question can also be posed to study the trends of the various emotions over the lifespan of a particular support ticket.

Furthermore, the ultimate goal is to develop a model that can reliably predict if, and when, a support ticket may be escalated. Given the results described in this paper, we believe that a model \textit{can} be developed; however, at this time we do not necessarily know \textit{how} to develop such a model. That being said we propose an additional query to investigate \textit{how} the extracted sentiment and emotions from a support ticket can be utilized in assisting the prediction and reaction to potential escalations. This subsequent research will harness both sentiment analysis and machine learning tools to provide insight as to how such a model may be developed and used in industry standard tools in order to provide escalation prediction to industry support analysts.
\section{Threats to Validity}
\label{sec:threats}
The biggest threats to the validity of the conclusions drawn from the data are the various assumptions and simplifications that were made to be able to conduct the analysis with the data that could be gathered. At each place such an assumption or simplification was made, it was described. On the other hand, since the data were mined well after the fact, there is no chance that these data could have been biased by a desire to support any particular hypothesis. In addition, the sentiment analysis tools used were developed using the context of  various data from social media platforms as input and therefore the results against technical support tickets may not be applicable. However, given that the tools were systematically, consistently, and equally applied to all data, the results should be consistent across the data, even if a particular tool was unable to comprehend any technical jargon. Furthermore, we have compared the results of three unique tools to overrule the bias of any one tool. Finally, as with any other case study, the results are about the particular case and cannot be generalized to other situations. So, we encourage others to do similar data mining studies on other data.

\section{Conclusion}
\label{sec:conclusions}
A support ticket escalation can be a costly experience for software organizations. Our research described in this paper attempts to identify indicators of when a particular support ticket may be escalated using three sentiment analysis tools: NLTK, SentiStrength, and Watson NLU. Using these tools we analyzed a very large support ticket database with assistance from our industrial partner, IBM, and found that there are notable differences between the sentiment in the escalated support tickets and the sentiment in the non-es\-ca\-la\-ted support tickets. These preliminary results provide us with confidence in our belief that we will be able to develop a model, based on sentiment analysis, which provides the ability for support analysts to focus on support tickets that have the potential to enter the escalation process.

\bibliographystyle{splncs}
\bibliography{main}

\end{document}